\documentclass[conference]{IEEEtran}
\IEEEoverridecommandlockouts
\usepackage{cite}
\usepackage{amsmath,amssymb,amsfonts}
\usepackage{graphicx}
\usepackage{textcomp}
\usepackage{xcolor}
\usepackage{fancyhdr} 
\usepackage{adjustbox}

\usepackage{color}
\usepackage{multirow}
\usepackage{booktabs}
\usepackage{caption}   
\usepackage{subfigure} 
\usepackage{comment}
\usepackage{mathtools}
\usepackage{mathrsfs}
\usepackage{algorithm}
\usepackage{algpseudocode}
\usepackage{amsmath}
\usepackage{siunitx}
\usepackage{url}

\usepackage{subcaption}
\usepackage{enumitem}
\usepackage{multicol}
\usepackage{multirow}
\usepackage{makecell}
\usepackage{bm}
\usepackage{xcolor}
\usepackage{colortbl}
\definecolor{my_gray}{rgb}{0.9216, 0.9216, 0.9216}
\definecolor{my_green}{rgb}{0.078, 0.686, 0.384}

\def\BibTeX{{\rm B\kern-.05em{\sc i\kern-.025em b}\kern-.08em
    T\kern-.1667em\lower.7ex\hbox{E}\kern-.125emX}}

\DeclareRobustCommand*{\IEEEauthorrefmark}[1]{%
    \raisebox{0pt}[0pt][0pt]{\textsuperscript{\footnotesize\ensuremath{#1}}}}
    
\begin{document}

\title{DISCO: A Hierarchical Disentangled Cognitive Diagnosis Framework for Interpretable \\ Job Recommendation}



\author{
\IEEEauthorblockN{
Xiaoshan Yu\IEEEauthorrefmark{1},
Chuan Qin\IEEEauthorrefmark{2,*}\thanks{$^{*}$Chuan Qin and Haiping Ma are corresponding authors.},
Qi Zhang\IEEEauthorrefmark{3},
Chen Zhu\IEEEauthorrefmark{4},
Haiping Ma\IEEEauthorrefmark{5,*},
Xingyi Zhang\IEEEauthorrefmark{6},
Hengshu Zhu\IEEEauthorrefmark{7}
}
\IEEEauthorblockA{
\IEEEauthorrefmark{1}School of Artificial Intelligence, Anhui University, China}

\IEEEauthorblockA{\IEEEauthorrefmark{2}Artifcial Intelligence Thrust, The Hong Kong University of Science and Technology (Guangzhou), China}

\IEEEauthorblockA{
\IEEEauthorrefmark{3}Shanghai Artificial Intelligence Laboratory, China}

\IEEEauthorblockA{\IEEEauthorrefmark{4}University of Science and Technology of China, School of Management, China}

\IEEEauthorblockA{
\IEEEauthorrefmark{5}Information Materials and Intelligent
Sensing Laboratory of Anhui Province, Institutes of \\ Physical Science and Information Technology, Anhui University, China}

\IEEEauthorblockA{\IEEEauthorrefmark{6}School of Computer Science and Technology, Anhui University, China}

\IEEEauthorblockA{\IEEEauthorrefmark{7}Computer Network Information Center, Chinese Academy of Sciences, China \\ \{yxsleo, chuanqin0426, zhangqi.fqz, zc3930155, xyzhanghust, zhuhengshu\}@gmail.com, hpma@ahu.edu.cn}

}






\maketitle



\begin{abstract}


The rapid development of online recruitment platforms has created unprecedented opportunities for job seekers while concurrently posing the significant challenge of quickly and accurately pinpointing positions that align with their skills and preferences. Job recommendation systems have significantly alleviated the extensive search burden for job seekers by optimizing user engagement metrics, such as clicks and applications, thus achieving notable success. In recent years, a substantial amount of research has been devoted to developing effective job recommendation models, primarily focusing on text-matching based and behavior modeling based methods. While these approaches have realized impressive outcomes, it is imperative to note that research on the explainability of recruitment recommendations remains profoundly unexplored. To this end, in this paper, we propose \textbf{DISCO}, a hierarchical \underline{\textbf{Dis}}entanglement based \underline{\textbf{Co}}gnitive diagnosis framework, aimed at flexibly accommodating the underlying representation learning model for effective and interpretable job recommendations. Specifically, we first design a hierarchical representation disentangling module to explicitly mine the hierarchical skill-related factors implied in hidden representations of job seekers and jobs. Subsequently, we propose level-aware association modeling to enhance information communication and robust representation learning both inter- and intra-level, which consists of the inter-level knowledge influence module and the level-wise contrastive learning. Finally, we devise an interaction diagnosis module incorporating a neural diagnosis function for effectively modeling the multi-level recruitment interaction process between job seekers and jobs, which introduces the cognitive measurement theory. Extensive experiments on two real-world recruitment recommendation datasets and an educational recommendation dataset clearly demonstrate the effectiveness and interpretability of our proposed DISCO framework. Our codes are available at \url{https://github.com/LabyrinthineLeo/DISCO}.





\end{abstract}



\section{introduction}

\begin{figure}[!t] 
    \centering 
    \includegraphics[width=0.99\linewidth]{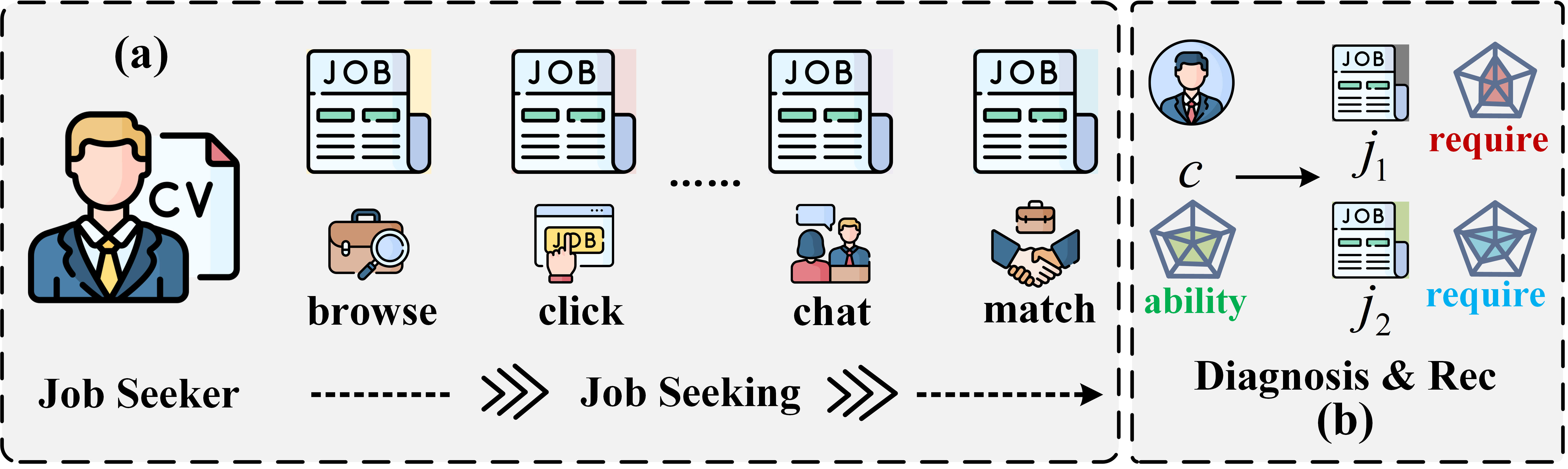} 
    \vspace{-2mm}
    \caption{An illustrative example of the recruitment diagnosis and job recommendation process.}
    \label{fig.gcd} 
    \vspace{-4mm}
\end{figure}


With the rapid development of Internet technology, various online recruitment platforms have emerged and become prevalent, such as LinkedIn and Glassdoor, which have significantly revolutionized the job-seeking process by establishing a digital bridge between job seekers and potential employers worldwide~\cite{qin2023comprehensive,yao2022knowledge}. Under this online paradigm of recruitment, it is crucial to develop a trustworthy job recommender system capable of accurately suggesting positions that align with job seekers' preferences and capabilities, which can assist them in efficiently finding the most suitable and credible positions~\cite{zha2023career}.
In the literature, there are many recent studies dominated by text-matching based methods~\cite{APJFNN,PJFNN}, these job recommender systems parsed through vast textual data from resumes and job postings to identify matches based on textual similarity~\cite{fang2023recruitpro,qin2022towards,qin2023automatic}, offering tailored job suggestions to users. In recent years, another category of interaction behavior based job recommendation methods~\cite{DPGNN,SHPJF,jiang2024towards,yao2021interactive,chen2024job} has received increasing research attention, which mainly explores users' personalized preferences and intentions by modeling the interaction behaviors between job seekers and recruiters. For example, DPGNN~\cite{DPGNN} introduces a dual-perspective graph representation learning framework, specifically designed to model the directed interactions (such as job application and resume review) between job seekers and recruiters. 
However, previous approaches have markedly enhanced the recommendation performance primarily by using end-to-end network models, there remains a notable deficiency in exploring the interpretability, particularly regarding the explainable abilities of job seekers and the variances in job requirements. As shown in Figure~1, the job seeker $c$ is presented with two positions (i.e., $j_1$ and $j_2$) by the recommender system that appears indistinguishable in terms of skills (indeed the two are extremely different with respect to the competencies matched to this user), which is a common issue in traditional job recommendation approaches. The lack of reasons for recommendations often makes it difficult for users to choose from recommendations, and they are more susceptible to position bias, ignoring the match between personal abilities and the needs of the position. This kind of recommender system may lead job seekers to pursue positions misaligned with their abilities or career aspirations, hindering their job search success.
To this end, in this paper, we propose a hierarchical \underline{\textbf{Dis}}entanglement based \underline{\textbf{Co}}gnitive diagnosis framework (DISCO), aimed at flexibly accommodating
the underlying representation learning model for effective and interpretable job recommendations. DISCO re-examines the recruitment recommendation process from a hierarchical disentanglement based cognitive diagnosis perspective, i.e., by viewing the process of job seekers interacting with the recruiter of the positions in the job search process as the process of exercise-solving by the learner~\cite{yang2023evolutionary,yang2024evolutionary,yang2024endowing,ReliCD}. Such a modeling approach facilitates the evaluation of the diverse skill demands of various job positions within the labor market and the current competitive standing of candidates. As illustrated in Figure~1, the competency level of candidate $c$ is explicitly depicted, alongside the skill requirement of jobs $j_1$ and $j_2$. This enhanced clarity enables candidates to better understand the differences in skill demands between various positions, allowing them to align their applications more precisely with jobs that match their capabilities and professional aspirations.


However, this task is non-trivial and presents several crucial technical challenges: (1)~\textit{How to map representations of users and jobs to specific skill dimensions to obtain interpretable content with tangible meanings?} Indeed, existing text-based and behavioral modeling job recommendation methods are primarily designed to learn effective representation vectors, which are high-order and abstract, failing to provide employers and job seekers with intuitive information, such as the degree of specificity of a job's skill requirements. 
(2)~\textit{How to hierarchize the competency level of candidates and the degree of skill required for jobs?} In recruitment interactions, the skills required for a job are usually multi-granular, and similarly, the level of competence possessed by job seekers is often hierarchical. Modeling these aspects in a coarse-grained manner fails to effectively capture their intrinsic characteristics. 
(3)~\textit{How to mitigate instability caused by interaction bias during the job search process?} In real-world scenarios, the interaction between job seekers and jobs during the job search process can be influenced by various biases, such as cognitive bias and popularity bias, which often leads to unstable profiling of the job seekers' abilities and job requirements.



To address these challenges, 
we initially developed a hierarchical representation disentangling module to effectively extract and clarify the hierarchical skill-related factors embedded in the hidden representations of job seekers and jobs. Subsequently, we design a level-aware self-attention network to explore the intrinsic associations between inter-level skill prototypes. Following that, a noise perturbation based level-wise contrastive module is proposed to enhance the robust representation learning. Finally, we devise an interaction diagnosis module is introduced that integrates a neural diagnosis function, aimed at effectively capturing the multi-level recruitment interaction process between job seekers and jobs. This module incorporates the cognitive measurement theory to enhance its explainability. Extensive experiments on two real-world recruitment recommendation datasets and an educational recommendation dataset clearly demonstrate the effectiveness and interpretability of the proposed DISCO framework in the job recommendation task.

\section{related work}

\subsection{Job Recommendation}
With the burgeoning of online recruitment platforms, job recommendation~\cite{sun2021market,sun2024large,sun2021discerning,shen2023exploiting} has emerged as a pivotal task and garnered extensive research attention, attributable to its potential to accurately match job seekers with suitable job positions. Existing job recommendation approaches are mainly classified into two categories, respectively, text-matching based methods and interaction behavior based methods. The first category of methods~\cite{PJFNN,APJFNN,PJFFF} focuses on matching the textual content of the applicants' resumes and job descriptions by utilizing text-matching strategies or text enhancement techniques~\cite{yang2024hybrid,chu2023architecture}, so as to predict the suitability of both. For example, APJFNN~\cite{APJFNN} employs a RNN to acquire word-level semantic representations and designs a hierarchical ability-aware attention strategy to measure the matching degree. Another category of methods~\cite{DPGNN,SHPJF} mainly explores users' personalized preferences and intentions by modeling the interaction behaviors between person (job seekers) and jobs (recruiters). For instance, SHPJF~\cite{SHPJF} explicitly models the users' search histories in addition to learning semantic information from text content for comprehensive mining of underlying job intents. Despite the impressive outcomes achieved by these methods, a significant shortfall is evident in the interpretable exploration of matching job seekers with job positions.





\vspace{-3mm}
\subsection{Cognitive Diagnosis}
Cognitive diagnosis (CD)~\cite{de2004higher,yang2024evolutionary_CD} is a classical methodology for assessing ability in educational psychology, aimed at portraying learners' proficiency profile by analyzing their learning behaviors~\cite{ma2024hd,RIGL}. Over the past decades, numerous effective CD models have been developed. Within these, traditional psychometric-based CD approaches hold a crucial role, being designed on the basis of psychological theories to depict student knowledge state through latent factors~\cite{de2004higher,DINA}. For instance, The Deterministic Inputs, Noisy And gate~(DINA)~\cite{DINA} model characterizes each student with a binary vector that indicates mastery of the knowledge concepts associated with the exercises, requiring all relevant skills for the highest positive response probability. In recent years, the swift development of deep learning has propelled neural network (NN)-based CD approaches~\cite{NeuralCD,RDGT,DGCD} to the forefront. These methods effectively diagnose learners' mastery attributes by incorporating neural networks to model complex interactions among learning elements (e.g., students, exercises, and knowledge concepts). For example, NeuralCD~\cite{NeuralCD}, a notable neural CDM, employs multidimensional parameters for detailed depiction of students' knowledge level and exercise attributes, and incorporates MLP to model complex interactions between students and exercises. RDGT~\cite{RDGT} proposes an effective group cognitive diagnosis method by designing a relation-guided dual-side graph transformer model to mine potential associations both between learners and between exercises. However, a gap remains in the skillful application of cognitive diagnostics to effectively model the job recommendation task.

\subsection{Disentangled Learning }
Disentangled learning aims to identify and disentangle the underlying explanatory factors of the observed complicated data, enhancing the efficiency and interpretability of the model in the learning process~\cite{DRL_Survey,DRL_GAN}. Initially, disentangled learning received extensive research attention in the field of computer vision due to its effectiveness~\cite{DRL_GAN}. Recently, a variety of approaches have introduced disentangled learning to model graph-structured data~\cite{DisenGCN,DGCL}. For instance, DGCL~\cite{DGCL} employs contrastive learning to uncover latent factors within the graph and subsequently extracts disentangled representations of the graph. DisenGCN~\cite{DisenGCN} proposes a unique neighborhood routing mechanism for disentangling node representation in graph networks, enabling dynamic identification of latent factors and improved performance in complex data scenarios. Moreover, learning disentangled representations of user latent intents from interaction feedback has been a popular topic in the recommendation domain~\cite{MacridVAE,wang2020setrank,shen2024handling}. For example, MacridVAE~\cite{MacridVAE} proposes a macro-micro disentangled variational auto-encoder to learn disentangled representations based on user behavior across multiple geometric spaces. However, learning disentangled competency representations of applicants in the cognitive diagnostic perspective is unexplored.

\begin{figure*}[!t] 
	\centering 
        \vspace{-2mm}
	\includegraphics[scale=0.97]{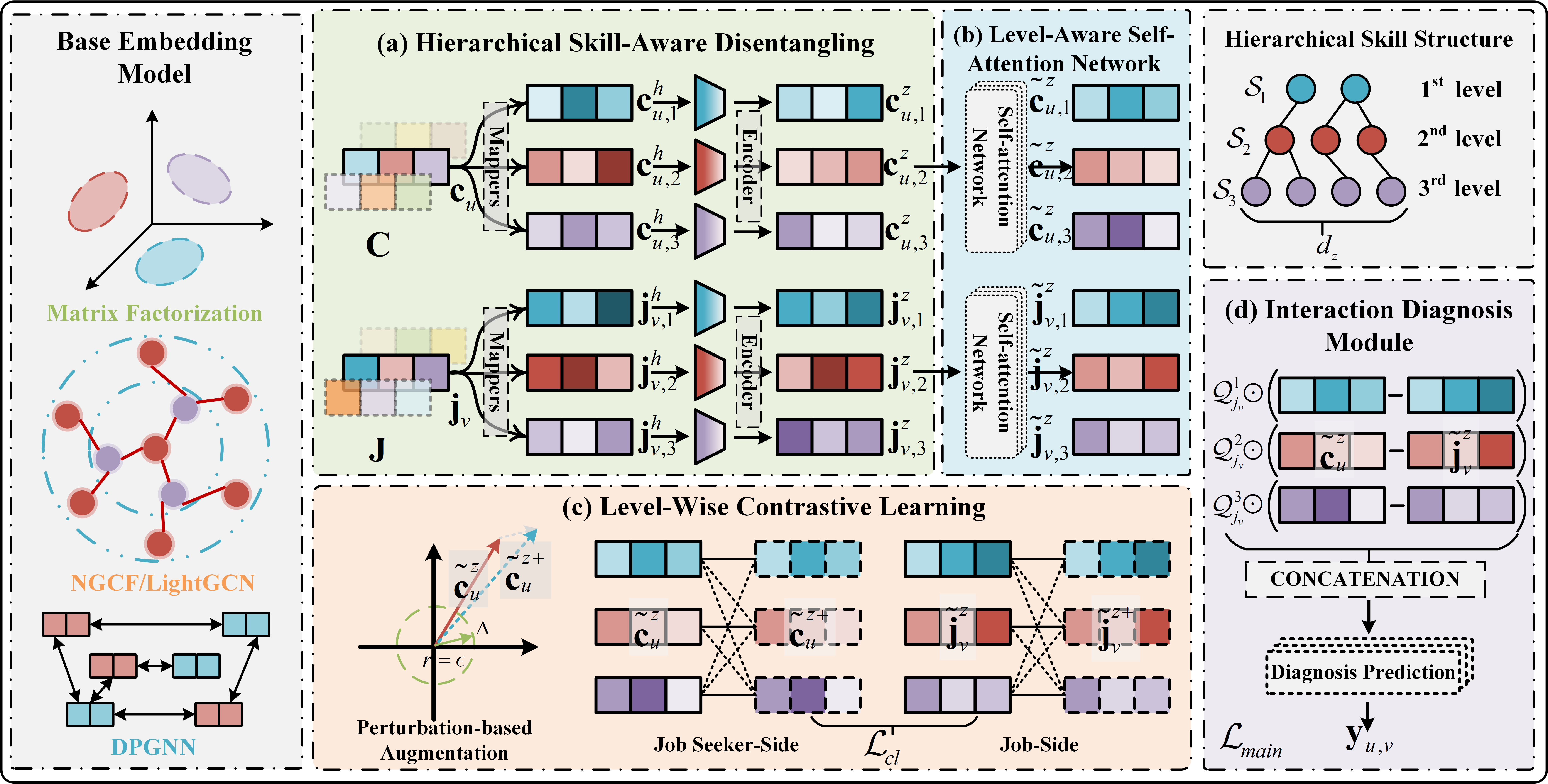} 
    \caption{The overview architecture of our proposed DISCO framework.}
        \label{fig.framework} 
        \vspace{-4mm}
\end{figure*}

\section{Preliminaries}

\subsection{Problem Formulation}

In this section, we introduce the problem definition of job recommendation. Let $\mathcal{C} = \{c_1, c_2, \ldots, c_N\}$ be the set of $N$ job seekers, $\mathcal{J} = \{j_1, j_2, \ldots, j_M\}$ be the set of $M$ jobs, $\mathcal{S} = \cup_{l=1}^{L}\mathcal{S}_l$ be the set of $K$ skills with $L$ different granularity levels, where $\mathcal{S}_L$ is the atomic skill level and $K = \sum_{l=1}^L|\mathcal{S}_l|$. Each job seeker and job are associated with textual documents describing the resumes and the job requirements. The relationship between jobs and skills is represented by a $Q$-matrix $\mathcal{Q} = \{q_{uv}\}^{M\times K}$, where $q_{uv} = 1$ if job $j_u$ requires skill $s_v$ and $0$ otherwise. Besides, the interaction matrix between the job seekers and jobs is denoted as $\mathcal{R} = \{r_{uv}\}^{N\times M}$, where $r_{uv} \in \{0,1,2,3\}$ corresponds to the four kinds of interaction behaviors between the candidate $c_u$ and the job $j_v$, i.e., \textit{Browse}, \textit{Click}, \textit{Chat} and \textit{Match}, respectively, and they reflect different levels of matching. Particularly, $r_{uv} = 3$ means both the job seeker and the recruiter are satisfied with each other and this pair is matched.

In this paper, our goal is to recommend the appropriate jobs to job seekers. To achieve this, we aim to predict the compatibility between jobs and candidates by learning a matching function $f(c_u,j_v)$ based on the interaction records $\mathcal{R}$, the relationship matrix between jobs and skills $\mathcal{Q}$, and the documents describing the resume and job requirements, and then realize the top-$K$ job recommendation based on the predicted degree of matching between candidates and jobs.


\subsection{Base Embedding Model}

The goal of DISCO is primarily to efficiently and interpretably model the interaction patterns between users and jobs, thus enabling a flexible framework applicable to existing representation learning recommendation models. In this section, we mainly introduce the base embedding model to output embedding representations of users and jobs. Specifically, the base embedding model $\mathcal{M}(\mathcal{C}, \mathcal{J}, \mathcal{R})$ aims to encode job seekers $\mathcal{C}$ and jobs $\mathcal{J}$ with $d$-dimensional trainable matrices $\textbf{C} \in \mathbb{R}^{N\times d}$ and $\textbf{J} \in \mathbb{R}^{M \times d}$, respectively. Here $N$ and $M$ are the numbers of job seekers and jobs in the recruitment interaction records, respectively, and $d$ is the embedding dimension. Notably, the process of acquiring embedding representations: $\mathcal{M(\mathcal{C}, \mathcal{J},\mathcal{R})} \rightarrow (\textbf{C}, \textbf{J})$, varies across different base models (as shown in the left part of Figure~2), with the essential purpose being to learn effective high-dimensional characterizations. Specifically, we deploy MF~\cite{MF} to consider collaborative information, NGCF/LightGCN~\cite{NGCF,LightGCN} to extract higher-order connectivity, and DPGNN~\cite{DPGNN} to embed textual content.
Here, we take the example of NGCF~\cite{NGCF}, which utilizes the user-item interaction graph to propagate high-order information and thus achieve embedding learning:
\begin{equation}
    \footnotesize
    \setlength{\abovedisplayskip}{5pt}
    \setlength{\belowdisplayskip}{3pt}
    \label{NGCF}
    \left\{
    \begin{aligned}
        \textbf{c}_u^{(k+1)} &= \sigma \Big(\textbf{W}_1\textbf{c}_u^{(k)} + \sum_{v\in\mathcal{N}_u}\frac{1}{\sqrt{|\mathcal{N}_u||\mathcal{N}_v|}}(\textbf{W}_1\textbf{j}_v^{(k)}+\textbf{W}_2(\textbf{j}_v^{(k)} \odot \textbf{c}_u^{(k)} )) \Big), \\
        \textbf{j}_v^{(k+1)} &= \sigma \Big(\textbf{W}_1\textbf{j}_v^{(k)} + \sum_{u\in\mathcal{N}_v}\frac{1}{\sqrt{|\mathcal{N}_u||\mathcal{N}_v|}}(\textbf{W}_1\textbf{c}_u^{(k)}+\textbf{W}_2(\textbf{c}_u^{(k)} \odot \textbf{j}_v^{(k)} )) \Big),
    \end{aligned}
    \right.
\end{equation}
where $\textbf{c}_u^{(k)}$ and $\textbf{j}_v^{(k)}$ are the refined embedding of user $c_u$ and item $j_v$ after $k$ layers propagation, respectively, $\sigma$ denotes the nonlinear activation function, $\odot$ refers to the element-wise multiply operator, $\mathcal{N}_u$ and $\mathcal{N}_v$ respectively represent the set of items interacted by user $c_u$ and the set of users interacted by item $j_v$, and $\textbf{W}_1$ and $\textbf{W}_2$ are trainable weight matrix to conduct feature
transformation in each layer. Finally, the representations of all the layers are concatenated to obtain the output embeddings, i.e., $\textbf{c}_u = \underset{l=0}{\stackrel{L}{||}}\!\textbf{c}_u^{(l)}$ and $\textbf{j}_v = \underset{l=0}{\stackrel{L}{||}}\!\textbf{j}_v^{(l)}$, where $\textbf{c}_u \in \textbf{C}$, $\textbf{j}_v \in \textbf{J}$, and $||$ denotes the concatenation operation.

\section{methodology}


In this section, we present the DISCO framework in detail. As illustrated in Figure~2, the architecture of DISCO consists of four main components, including the hierarchical skill-aware representation disentangling, the level-aware self-attention network, level-wise contrastive learning, and the interaction diagnosis module.

\subsection{Hierarchical Representation Disentangling} 

\subsubsection{Modeling Latent Skill Factors.}


In the job recommendation task, the main goal is to predict the matching degree between a job seeker and a job given their observed interactions. During the real recruitment process, when a candidate interacts with a job, the outcome is often influenced by skill-related factors, such as the candidate's level of skill mastery and the difficulty of the skills required by the job~\cite{DuerQuiz}. To capture these skill factors, we formally define our prediction objective about job seeker-job matching as follows:
\begin{equation}
    \label{log_pred}
    \begin{split}
        p_\theta(y|X) = \mathbb{E}_{p_\theta(z_c|X)p_\theta(z_j|X)} [p_{\theta}(y|X, z_c, z_j)],
    \end{split}
\end{equation}
where $X=(c_u,j_v,s_{j_v})$ denotes the candidate, the job, and the skills involved in the job description, respectively, $y$ is the learned matching score. Hence the optimization objective of the predictive model is as follows:
\begin{equation}
    \label{log_pred}
    \begin{split}
        \theta^* &= \underset{\theta}{\textit{arg max}} \sum_{i=1}^{|\mathcal{R}|} log\;p_\theta(y_i|X_i), \\
        &= \underset{\theta}{\textit{arg min}} \sum_{i=1}^{|\mathcal{R}|} -log\;\mathbb{E}_{p_\theta(z_c|X_i)p_\theta(z_j|X_i)} [p_\theta(y_i|X_i, z_c, z_j)],
    \end{split}
\end{equation}
where $|\mathcal{R}|$ denotes the number of observed samples, and $z_c$ and $z_j$ are the skill factors for candidates and jobs, respectively. We use $g(\cdot)$ to denote the prediction function over the encoded skill factors. Referring to previous work~\cite{DCCF}, we approximate our prediction objective as follows:
\begin{equation}
    \small
    \label{log_pred}
    \begin{split}
    \mathbb{E}_{p_\theta(z_c|X_i)p_\theta(z_j|X_i)} [g(z_c,z_j)] \approx g(\mathbb{E}_{p_\theta(z_c|X_i)}[z_c], \mathbb{E}_{p_\theta(z_j|X_i)}[z_j]).
    \end{split}
\end{equation}
Considering the above inference, the approximation error can be effectively constrained within our prediction function $g(\cdot)$.

\subsubsection{Hierarchical Skill-Aware Disentangling.} 

In the recruitment recommendation process, skill-related factors~(e.g., skill proficiency and skill demand) are the primary determinants of the interaction outcome between the job seeker and the job position~\cite{sun2021market}. Explicitly disentangling these factors from the embedding representations of the candidates and the jobs is quite important, which facilitates understanding of the job seeker's mastery of the skill and the job's requirement of the skill, thus enhancing interpretability. Indeed, the skills involved in a job may be at different levels of granularity~\cite{DuerQuiz}, which tend to encompass relationships~(as shown in Figure~2). Towards this end, we propose to disentangle hierarchical skill characteristics from the high-dimensional and abstract representations of users and jobs that incorporate interaction information. Specifically, for the embeddings $\textbf{c}_u\in \textbf{C}$ and $\textbf{j}_v\in \textbf{J}$ of user $c_u$ and job $j_v$ obtained from the base model (as mentioned before), we first construct $L$-layer mappers to project them into different hierarchical skill spaces, respectively, as follows:
\begin{equation}
    \label{projector}
    \begin{split}
        \textbf{c}_{u,l}^{h} = \textbf{c}_u \textbf{W}_l^{c}, \; \textbf{j}_{v,l}^{h} = \textbf{j}_v \textbf{W}_l^{j}, \; 1\leq l\leq L,
    \end{split}
\end{equation}
where $\textbf{c}_{u,l}^{h}\in \mathbb{R}^{d_h}$ and $\textbf{j}_{v,l}^{h} \in \mathbb{R}^{d_h}$ denotes the mapped hidden representations of user $c_u$ and job $j_v$ at $l$-th skill layer, respectively, $\textbf{W}_l^{c},\textbf{W}_l^{j} \in \mathbb{R}^{d\times d_h}$ are the trainable matrices, and $d_h$ is the hidden dimension. Here, $L$ represents the number of skill layers at different levels of granularity, which is a fixed parameter that comes with the dataset~(as mentioned in Section~3.1). Subsequently, we build multi-level encoders for users and jobs respectively to learn $L$ ability prototypes $\{\textbf{c}_{u,l}^z\in \mathbb{R}^{d_z}\}_{l=1}^{L}$ and $L$ skill difficulty prototypes $\{\textbf{j}_{v,l}^z\in \mathbb{R}^{d_z}\}_{l=1}^{L}$ for the hierarchical skill space as follows: 
\begin{equation}
    \label{encoder}
    \begin{split}
        \textbf{c}_{u,l}^{z} = Encoder_l^c(\textbf{c}_{u,l}^{h}), \; \textbf{j}_{v,l}^{z} = Encoder_l^j(\textbf{j}_{v,l}^{h}), \; 1\leq l\leq L,
    \end{split}
\end{equation}
where $Encoder_l^c$ and $Encoder_l^j$ denote the $l$-th layer disentangled encoder for the users and jobs, respectively. Similar to previous work~\cite{DGCL}, a multilayer perceptron network (MLP) is used here. Notably, our goal is to disentangle the various levels of competencies possessed by job seekers as well as the distinct tiers of skill levels required by job positions. In particular, $d_z = |\mathcal{S}_L|$ denotes the number of atomic skills, thereby facilitating the explicit alignment of the dimensions of both representation vectors with the skill size.



\vspace{-2mm}
\subsection{Level-Aware Association Modeling}

\subsubsection{Inter-Level Knowledge Influence}

While it is desirable for the disentangled prototypes to effectively characterize users and jobs at different skill levels, in reality, these inter-level skill representations are often influenced by underlying knowledge. For example, a job seeker's mastery of the coarse-grained skills is usually affected by their proficiency in finer-grained skills, and conversely, the same principle applies. To further explore the intrinsic associations between inter-level skill prototypes, we design a level-aware self-attention network for enhancing learning as follows:
\begin{equation}
    \centering
    \label{Encoder_SelfAtt}
    \left\{
    \begin{aligned}
         \tilde{\textbf{c}}_{u,l}^z &= SelfAtt(\mathcal{Q}^c,\mathcal{K}^c,\mathcal{V}^c), \\
         \mathcal{Q}^c&=\textbf{c}_{u,l}^z, \mathcal{K}^c=\{\textbf{c}_{u,1}^z,\ldots,{\textbf{c}}_{u,L}^z\}, \mathcal{V}^c=\{\textbf{j}_{v,1}^z,\ldots,{\textbf{j}}_{v,L}^z\}, \\
         \tilde{\textbf{j}}_{v,l}^z &= SelfAtt(\mathcal{Q}^j,\mathcal{K}^j,\mathcal{V}^j), \\
         \mathcal{Q}^j&=\textbf{j}_{v,l}^z, \mathcal{K}^j=\{\textbf{j}_{v,1}^z,\ldots,{\textbf{j}}_{v,L}^z\}, \mathcal{V}^j=\{\textbf{j}_{v,1}^z,\ldots,{\textbf{j}}_{v,L}^z\}. \\
    \end{aligned}
    \right. 
\end{equation}
where $\tilde{\textbf{c}}_{u,l}^z, \tilde{\textbf{j}}_{v,l}^z \in \mathbb{R}^{d_z}$ are enhanced $l$-level skill aware representaitons incorporating correlated information,  $\mathcal{Q}^c$,$\mathcal{K}^c$,$\mathcal{V}^c$ and $\mathcal{Q}^j$,$\mathcal{K}^j$,$\mathcal{V}^j$ denote the query, key and value vectors for the user and job, respectively, and $SelfAtt(\cdot)$ indicate the Self-Attention module~\cite{SelfAtt}.   


\subsubsection{Level-Wise Contrastive Learning}

In real-world scenarios, the interaction between job seekers and jobs during the job search process can be influenced by various biases~\cite{qin2020enhanced,PJFNN}, such as cognitive bias and popularity bias. This often leads to unstable profiling of the job seekers' abilities and the job's requirements, particularly in the context of disentangled representations, resulting in sub-optimal performance as well as inaccurate interpretations. Taking inspiration from recent developments in contrastive learning, we attempt to enhance the robustness of skill-aware disentangled representations by exploring self-supervised signals. Different from previous work that constructs contrastive tasks from instance dimensions~\cite{DCCF}, we propose a level-wise contrastive learning for modeling between skill representations at different granularities. Specifically, we formalize the level-wise contrastive learning loss (from the job seeker side) as follows:
\begin{equation}
    \centering
    \label{CL_loss_1}
    \begin{split}
        \mathcal{L}^{C\!L}_c = \frac{1}{L}\sum_{u=1}^{|\mathcal{C}|} \sum_{l=1}^{L} -log\; p_\theta(c_u'|c_u,z_{c,l}),
    \end{split} 
\end{equation}
where $p_\theta(c_u'|c_u,z_{c,l})$ denotes the candidate ability contrastive learning subtask under $l$-th skill level, and $z_{c,l}$ is the $l$-th level latent skill factor of the job seeker (i.e., ability). We aim to learn the optimal $L$ ability prototypes which are able to maximize the expectation of $L$ subtasks, and the contrastive learning subtask for the $l$-level ability is defined as follows:
\begin{equation}
    \centering
    \label{CL_loss_2}
    \begin{split}
        p_\theta(c_u'|c_u,z_{c,l}) = \frac{exp(\phi(\tilde{\textbf{c}}_{u,l}^z,\tilde{\textbf{c}}_{u,l}^{z+})/\tau)}{\sum_{l'=1}^L exp(\phi(\tilde{\textbf{c}}_{u,l'}^z,\tilde{\textbf{c}}_{u,l'}^{z+})/\tau)},
    \end{split} 
\end{equation}
where $\tilde{\textbf{c}}_{u,l}^z$ and $\tilde{\textbf{c}}_{u,l}^{z+}$ are the user positive pair of the $l$-level ability, $\tau$ is a temperature parameter, $\phi(\cdot)$ denotes the similarity function, and we use the cosine similarity function here:
\begin{equation}
    \centering
    \label{CosSim}
    \begin{split}
        \phi(\tilde{\textbf{c}}_{u,l}^z,\tilde{\textbf{c}}_{u,l}^{z+}) = \frac{\tilde{\textbf{c}}_{u,l}^z{}^T \tilde{\textbf{c}}_{u,l}^{z+}}{\|\tilde{\textbf{c}}_{u,l}^z\|_2\|\tilde{\textbf{c}}_{u,l}^{z+}\|_2}.
    \end{split} 
\end{equation}

In particular, $\tilde{\textbf{c}}_{u,l}^{z+}$ here is the augmented $l$-level ability representation of the job seeker $c_u$. Inspired by~\cite{SimGCL}, we implement an efficient and effective skill-level augmentation by directly adding random noises to the ability representation as follows:
\begin{equation}
    \centering
    \label{CosSim}
    \begin{split}
        \tilde{\textbf{c}}_{u,l}^{z+} = \tilde{\textbf{c}}_{u,l}^z + \Delta_{u,l}',
    \end{split} 
\end{equation}
where the added noise vectors $\Delta_{u,l}'$ is subject to $\|\Delta\|_2 = \epsilon$ and $\Delta = \overline\Delta \odot sign(\tilde{\textbf{c}}_{u,l}^z), \overline\Delta \in \mathbb{R}^{d_z} \sim U(0,1)$, and $sign(\cdot)$ denotes the sign function and $U(0,1)$ represents the uniform distribution. The first constraint regulates the magnitude of $\Delta$, which is numerically akin to coordinates on a hypersphere with the radius $\epsilon$. The second constraint ensures that $\tilde{\textbf{c}}_{u,l}^z$ and $\Delta_{u,l}$ remain within the same hyperoctant to avoid significant deviations in $\tilde{\textbf{c}}_{u,l}^z$ when noise is added, thereby maintaining the validity of positive samples. Notably, the modeling process for the job side is the same as above.

\subsection{Interaction Diagnosis Module}

In this section, we present an interaction diagnosis module designed to model the nuanced interactions between job seekers and jobs at multiple granularity skill levels from a cognitive measurement perspective by introducing cognitive diagnosis theory~\cite{de2004higher,liu2023homogeneous,yang2023cognitive}.

\subsubsection{Neural Diagnosis Interaction}
Indeed, in cognitive diagnosis assessments~\cite{DINA,NeuralCD}, the key research focus is how to effectively measure a tester's ability level by modeling their ability representations against the difficulty representations of exercises across different knowledge concepts. In our DISCO framework, the already disentangled multi-level skill characteristics of job seekers and jobs, which are explicitly mapped to skill dimensions, facilitate the interaction modeling using diagnosis functions. Similar to~\cite{NeuralCD}, we adopt the neural diagnosis function in our framework. It can seamlessly integrate with non-linear neural network layers, and its capability to model high-dimensional interactive elements enables the acquisition of extensive knowledge and the presentation of interpretable information. It is formalized as follows (taking as an example the disentangled $l$-level skill representations of $c_u$ and $j_v$): 
\begin{equation}
    \label{InterFun}
    \begin{split}
         \mathcal{T}(\tilde{\textbf{c}}_{u,l}^z,\; \tilde{\textbf{j}}_{v,l}^z) &= \mathcal{Q}_{j_v}^l \odot (\sigma(\tilde{\textbf{c}}_{u,l}^z) - \sigma(\tilde{\textbf{j}}_{v,l}^z)),
    \end{split}
\end{equation}
where the $\tilde{\textbf{c}}_{u,l}^z \in \mathbb{R}^{d_z}$ and $\tilde{\textbf{j}}_{v,l}^z \in \mathbb{R}^{d_z}$ denote the $l$-level skill representations of the job seeker $c_u$ and the job $j_v$, respectively (mentioned above), $\sigma(\cdot)$ is the activation function (here is the Sigmoid function), $\odot$ refers to the element-wise multiply operator, and $\mathcal{Q}_{j_v}^l$ indicates the skill attribute corresponding to $j_v$ originates from the $Q$-matrix $\mathcal{Q}$. Thus, we obtain the matching distance in the $l$-level skill space between $c_u$ and $j_v$: $\textbf{h}_{u,v}^l = \mathcal{T}(\tilde{\textbf{c}}_{u,l}^z,\; \tilde{\textbf{j}}_{v,l}^z) \in \mathbb{R}^{d_z}$, which reflects the degree of matching between the two at the $l$-th skill level from a diagnosis perspective.

\subsubsection{Hierarchical Diagnosis Prediction}
To further aggregate the hierarchical competency matching distances between candidates and jobs and assess them comprehensively, we conduct a hierarchical diagnostic prediction. Specifically, we first concatenate the obtained L-layer matching distance representations as follows:
\begin{equation}
    \label{InterLayer}
    \begin{split}
         \textbf{h}_{u,v} = \underset{l=1}{\stackrel{L}{||}}\!\textbf{h}_{u,v}^l,
    \end{split}
\end{equation}
where $||$ denotes the concatenation operation, and $\textbf{h}_{u,v} \in \mathbb{R}^{L\cdot d_z}$ is the aggregated interaction vector, which serves to comprehensively account for the hierarchical effects of different skill levels. Subsequently, we utilize the full connection layers to model the high-order interaction features, as follows: 
\begin{equation}
    \label{InterLayer}
    \left\{
    \begin{aligned}
         \textbf{h}_{u,v}' &= \sigma(\textbf{h}_{u,v}\textbf{W}_1 \;+\; \textbf{b}_1),\\
         \textbf{y}_{u,v} &= \sigma(\textbf{h}_{u,v}'\textbf{W}_2 \;+\; \textbf{b}_2),
    \end{aligned}
    \right.
\end{equation}
where $\textbf{W}_1\in \mathbb{R}^{d_z\times d_h}, \textbf{b}_1 \in \mathbb{R}^{d_h}, \textbf{W}_2\in \mathbb{R}^{d_h\times d_{cls}}$ and $\textbf{b}_2 \in \mathbb{R}^{d_{cls}}$ are trainable parameters, $d_h$ and $d_{cls}$ are the hidden dimensions and the number of interaction categories, and $\textbf{y}_{u,v}$ denotes the predicted probabilities of the different interaction categories between job seek $c_u$ and job $j_v$.

\subsubsection{Loss Function}
In the training phase, we update the model parameters mainly by predicting the job seeker-job interaction categories. Specifically, for each interaction record $(c_u, j_v, r_{uv})$, we utilize the multi-class cross-entropy loss function for the prediction: 
\begin{equation}
    \label{Loss_main}
    \begin{split}
        \mathcal{L}_{main} = - \frac{1}{|\mathcal{R}|}\sum\limits_{(c_u,j_v, \textbf{r}_{u,v}) \in \mathcal{R}} \sum_{i=1}^{d_{cls}} \textbf{r}_{u,v}^i\; log\; \textbf{y}_{u,v}^i \;,
    \end{split}
\end{equation}
where $d_{cls}$ is the number of interaction categories. It is worth noting that $\textbf{r}_{u,v} \in \{0,1\}^{d_{cls}}$ is a category-dependent one-hot vector originating from $r_{uv}$. To optimize the self-supervised subtask, we construct the complete contrastive learning loss as follows:
\begin{equation}
    \label{Loss_cl}
    \begin{split}
        \mathcal{L}_{cl} = \mathcal{L}_c^{CL} + \mathcal{L}_j^{CL}.
    \end{split}
\end{equation}
Finally, we obtain the complete optimization objective function by summing the loss functions of the above two objectives:
\begin{equation}
    \label{Loss_all}
    \begin{split}
        \mathcal{L} = \mathcal{L}_{main} + \lambda \cdot \mathcal{L}_{cl},
    \end{split}
\end{equation}
where $\lambda$ is the weight coefficient to control the influence of contrastive signals. We can then train the whole model and optimize the model parameters utilizing gradient descent.

\begin{table}[!t]
    \centering
    \label{table.dataset_statistic}
    \caption{Statistics of all experimental datasets.}
    \vspace{-1mm}
    \resizebox{0.95\linewidth}{!}{ 
        \begin{tabular}{lrrr}
        \toprule
        Statistics & Technology & Service & Edu-Rec\\
        \midrule
        \#Candidates & 4,726 & 10,022 & 61,567  \\
        \#Items   & 34,962 & 23,866 & 20,828 \\
        \#Skills & 986 & 3,241 & 384 \\
        \#Interactions & 616,504 & 866,065 & 2,200,731 \\
        Avg. interactions per user & 130.45 & 86.41 & 35.74 \\
        Avg. skills per job/item & 14.76 & 22.94 & 4.38 \\
        \bottomrule
        \end{tabular}
    }
    \vspace{-5mm}
\end{table}

\section{experiments}

In this section, we conduct extensive experiments on three real-world datasets to validate the effectiveness and interpretability of the proposed DISCO framework.

\begin{table*}[!t]
    \setlength{\tabcolsep}{1.5pt}
    \renewcommand{\arraystretch}{1.36}
    \centering
    \label{table.Result_Table_1}
    \caption{Performance of DISCO embedded in four base models and baselines on two recruitment recommendation datasets on job recommendation. “$*$” denotes the statistically significant improvement of DISCO model compared to the best baseline method (i.e., two-sided t-test with \textit{p}$<$0.05). $\bold{Bold}$: the best, $\underline{\rm Underline}$: the runner-up.}
    \vspace{-1mm}
    \resizebox{0.97\linewidth}{!}{ 
        \Huge
        \begin{tabular}{cc|c|c|c|c|c||c|c|c|c|c}
            \toprule
            \multicolumn{2}{c||}{\textbf{Datasets}}  & \multicolumn{5}{c||}{\textbf{Technology}} & \multicolumn{5}{c}{\textbf{Service}} \\
            \midrule
            \makecell[c]{\textbf{Base Model}} & \multicolumn{1}{|c||}{\textbf{Method}}  & \textbf{AUC} & \textbf{HR@5} & \textbf{NDCG@5} & \textbf{HR@10} & \textbf{NDCG@10}  & \textbf{AUC} & \textbf{HR@5} & \textbf{NDCG@5} & \textbf{HR@10} & \textbf{NDCG@10} \\
            \midrule
            \multirow{5}{*}{MF} & \multicolumn{1}{|c||}{Normal} &  $0.6755_{\pm 0.0014}$ & $0.2518_{\pm 0.0135}$ & $0.2394_{\pm 0.0084}$ & $0.5835_{\pm 0.0121}$ &  $0.3428_{\pm 0.0088}$ & $0.6479_{\pm 0.0011}$ & $0.4831_{\pm 0.0020}$ & 
            $0.3015_{\pm 0.0039}$ & $0.7511_{\pm 0.0053}$ & $0.3879_{\pm 0.0042}$ \\
            & \multicolumn{1}{|c||}{NCF} &  $\underline{0.6997_{\pm 0.0020}}$ &
            $\underline{0.4457_{\pm 0.0092}}$ & $0.3090_{\pm 0.0089}$ & $0.6261_{\pm 0.0065}$ &
            $0.3671_{\pm 0.0081}$ &
            $0.6742_{\pm 0.0058}$ & $\underline{0.6119_{\pm 0.0063}}$ & $0.4572_{\pm 0.0054}$ & $\underline{0.8306_{\pm 0.0042}}$ &  $0.5414_{\pm 0.0048}$ \\
            & \multicolumn{1}{|c||}{AutoInt} & $0.6940_{\pm 0.0025}$ & $0.4342_{\pm 0.0039}$ & $\underline{0.3094_{\pm 0.0028}}$ & $0.5995_{\pm 0.0055}$ &
            $0.3626_{\pm 0.0038}$ & $0.6629_{\pm 0.0073}$ & $0.5892_{\pm 0.0143}$ & $0.4088_{\pm 0.0284}$ & $0.8158_{\pm 0.0050}$ & $0.4826_{\pm 0.0315}$ \\
            & \multicolumn{1}{|c||}{FINAL} & $0.6978_{\pm 0.0005}$ &
            $0.4345_{\pm 0.0035}$ & $0.3081_{\pm 0.0055}$ & $\underline{0.6265_{\pm 0.0037}}$ & $\underline{0.3699_{\pm 0.0055}}$ & $\underline{0.6771_{\pm 0.0012}}$ &  $0.5861_{\pm 0.0081}$ & $\underline{0.4638_{\pm 0.0041}}$ & $0.8281_{\pm 0.0027}$ & $\underline{0.5427_{\pm 0.0024}}$ \\
            \cmidrule{2-12}
             & \multicolumn{1}{|c||}{\cellcolor{my_gray}DISCO} & \cellcolor{my_gray}$ \bm{0.7040^*_{\pm 0.0008}}$ & \cellcolor{my_gray}$\bm{0.4641^*_{\pm 0.0042}}$ & \cellcolor{my_gray}$\bm{0.3411^*_{\pm 0.0048}}$ & \cellcolor{my_gray}$\bm{0.6337^*_{\pm 0.0106}}$ &
             \cellcolor{my_gray}$\bm{0.3891^*_{\pm 0.0070}}$ & 
             \cellcolor{my_gray}$\bm{0.6892^*_{\pm 0.0055}}$ & \cellcolor{my_gray}$\bm{0.6527^*_{\pm 0.0017}}$ & \cellcolor{my_gray}$\bm{0.4701^*_{\pm 0.0037}}$ & \cellcolor{my_gray}$\bm{0.8414^*_{\pm 0.0015}}$ & \cellcolor{my_gray}$\bm{0.5516^*_{\pm 0.0034}}$ \\
            \midrule
            \multirow{5}{*}{NGCF} & \multicolumn{1}{|c||}{Normal} &  $0.7101_{\pm 0.0195}$ & $0.4705_{\pm 0.0268}$ & $0.3377_{\pm 0.0262}$ & $0.6144_{\pm 0.0472}$ & 
            $0.3840_{\pm 0.0205}$ & $0.6772_{\pm 0.0328}$ & $0.6523_{\pm 0.0144}$ & $0.4650_{\pm 0.0206}$ & $0.8258_{\pm 0.0203}$ & $0.5213_{\pm 0.0166}$ \\
            & \multicolumn{1}{|c||}{NCF} & $0.7245_{\pm 0.0025}$ & $0.5118_{\pm 0.0119}$ & $0.3601_{\pm 0.0079}$ & 
            $0.6978_{\pm 0.0101}$ & $0.4203_{\pm 0.0073}$ & $0.6856_{\pm 0.0021}$ &  $0.6547_{\pm 0.0025}$ & $0.4716_{\pm 0.0017}$ & $\underline{0.8375_{\pm 0.0011}}$ & $0.5326_{\pm 0.0012}$ \\
            & \multicolumn{1}{|c||}{AutoInt} & $\underline{0.7366_{\pm 0.0012}}$ & $\underline{0.5263_{\pm 0.0066}}$ & $0.3675_{\pm 0.0068}$ &
            $0.7113_{\pm 0.0084}$ & $\underline{0.4291_{\pm 0.0073}}$ & 
            $0.7179_{\pm 0.0033}$ &  $\underline{0.6559_{\pm 0.0096}}$ & $0.5029_{\pm 0.0070}$ & $0.8359_{\pm 0.0036}$ & $\underline{0.5563_{\pm 0.0103}}$ \\
            & \multicolumn{1}{|c||}{FINAL} & $0.7321_{\pm 0.0018}$ & $0.5239_{\pm 0.0063}$ & $\underline{0.3692_{\pm 0.0061}}$ &
            $\underline{0.7134_{\pm 0.0054}}$ & $0.4284_{\pm 0.0051}$ & $\underline{0.7231_{\pm 0.0023}}$ &  $0.6443_{\pm 0.0027}$ & $\underline{0.5110_{\pm 0.0066}}$ & $0.8291_{\pm 0.0008}$ & $0.5512_{\pm 0.0074}$ \\
            \cmidrule{2-12}
             & \multicolumn{1}{|c||}{\cellcolor{my_gray}DISCO} & \cellcolor{my_gray}$ \bm{0.7408^*_{\pm 0.0013}}$ & \cellcolor{my_gray}$\bm{0.5311^*_{\pm 0.0016}}$ & \cellcolor{my_gray}$\bm{0.3739^*_{\pm 0.0031}}$ & \cellcolor{my_gray}$\bm{0.7145^*_{\pm 0.0042}}$ & 
             \cellcolor{my_gray}$\bm{0.4338^*_{\pm 0.0041}}$ &  \cellcolor{my_gray}$\bm{0.7282^*_{\pm 0.0010}}$ & \cellcolor{my_gray}$\bm{0.6616^*_{\pm 0.0112}}$ & \cellcolor{my_gray}$\bm{0.5182^*_{\pm 0.0086}}$ & \cellcolor{my_gray}$\bm{0.8439^*_{\pm 0.0036}}$ & \cellcolor{my_gray}$\bm{0.5728^*_{\pm 0.0095}}$ \\
            \midrule
            \multirow{5}{*}{LightGCN} & \multicolumn{1}{|c||}{Normal} &  $0.7085_{\pm 0.0204}$ & $0.4741_{\pm 0.0056}$ & $0.3112_{\pm 0.0322}$ & $0.6364_{\pm 0.0304}$ &  $0.3637_{\pm 0.0089}$ & 
            $0.7051_{\pm 0.0178}$ & $0.6285_{\pm 0.0428}$ & $0.4273_{\pm 0.0462}$ & $0.8356_{\pm 0.0155}$ & $0.4943_{\pm 0.0374}$ \\
            & \multicolumn{1}{|c||}{NCF} & $0.7150_{\pm 0.0016}$ & $0.5007_{\pm 0.0022}$ & $0.3513_{\pm 0.0018}$ & $0.6945_{\pm 0.0029}$ &  
            $0.4106_{\pm 0.0023}$ & $0.7223_{\pm 0.0018}$ & $0.6661_{\pm 0.0041}$ & $0.5087_{\pm 0.0039}$ & $0.8229_{\pm 0.0009}$ & $0.5469_{\pm 0.0036}$ \\
            & \multicolumn{1}{|c||}{AutoInt} & $0.7124_{\pm 0.0008}$ & $\underline{0.5115_{\pm 0.0039}}$ & $0.3683_{\pm 0.0008}$ & $0.7023_{\pm 0.0051}$ & 
            $\underline{0.4218_{\pm 0.0015}}$ & $\underline{0.7279_{\pm 0.0030}}$ & $\underline{0.6717_{\pm 0.0099}}$ & $0.4905_{\pm 0.0051}$ & $0.8340_{\pm 0.0019}$ & $0.5513_{\pm 0.0032}$ \\
            & \multicolumn{1}{|c||}{FINAL} & $\underline{0.7159_{\pm 0.0029}}$ & $0.5068_{\pm 0.0015}$ & $\underline{0.3686_{\pm 0.0022}}$ & $\underline{0.7036_{\pm 0.0037}}$ &  
            $0.4127_{\pm 0.0024}$ & $0.7219_{\pm 0.0025}$ & $0.6338_{\pm 0.0056}$ & $\underline{0.5043_{\pm 0.0016}}$ & $\underline{0.8443_{\pm 0.0029}}$ & $\underline{0.5533_{\pm 0.0022}}$ \\
            \cmidrule{2-12}
             & \multicolumn{1}{|c||}{\cellcolor{my_gray}DISCO} & \cellcolor{my_gray}$ \bm{0.7262^*_{\pm 0.0033}}$ & \cellcolor{my_gray}$\bm{0.5221^*_{\pm 0.0020}}$ & \cellcolor{my_gray}$\bm{0.3722^*_{\pm 0.0042}}$ & \cellcolor{my_gray}$\bm{0.7070^*_{\pm 0.0075}}$ &  \cellcolor{my_gray}$\bm{0.4286^*_{\pm 0.0030}}$ & 
             \cellcolor{my_gray}$\bm{0.7321^*_{\pm 0.0053}}$ &  \cellcolor{my_gray}$\bm{0.6754^*_{\pm 0.0061}}$ & \cellcolor{my_gray}$\bm{0.5137^*_{\pm 0.0081}}$ & \cellcolor{my_gray}$\bm{0.8461^*_{\pm 0.0021}}$ & \cellcolor{my_gray}$\bm{0.5568^*_{\pm 0.0149}}$ \\
             \midrule
            \multirow{5}{*}{DPGNN} & \multicolumn{1}{|c||}{Normal} &  $0.7039_{\pm 0.0004}$ & $0.4986_{\pm 0.0025}$ & $0.3348_{\pm 0.0035}$ & $\underline{0.7018_{\pm 0.0019}}$ &  
            $0.4005_{\pm 0.0034}$ & $0.7089_{\pm 0.0017}$ & $0.6665_{\pm 0.0057}$ & $0.4664_{\pm 0.0061}$ & $0.8543_{\pm 0.0018}$ & $0.5274_{\pm 0.0044}$ \\
            & \multicolumn{1}{|c||}{NCF} & $\underline{0.7186_{\pm 0.0007}}$ & $0.5015_{\pm 0.0064}$ & $0.3542_{\pm 0.0049}$ &
            $0.6954_{\pm 0.0054}$ & $0.4168_{\pm 0.0047}$ & $0.7127_{\pm 0.0030}$ &  $\underline{0.6669_{\pm 0.0034}}$ & $\underline{0.4908_{\pm 0.0035}}$ & $0.8503_{\pm 0.0011}$ & $\underline{0.5479_{\pm 0.0028}}$ \\
            & \multicolumn{1}{|c||}{AutoInt} & $0.7144_{\pm 0.0020}$ & $0.5037_{\pm 0.0070}$ & $0.3506_{\pm 0.0089}$ & $0.6976_{\pm 0.0073}$ &  $0.4131_{\pm 0.0087}$ & 
            $0.7213_{\pm 0.0021}$ & $0.6533_{\pm 0.0026}$ & $0.4825_{\pm 0.0077}$ & $0.8514_{\pm 0.0010}$ & $0.5403_{\pm 0.0074}$ \\
            & \multicolumn{1}{|c||}{FINAL} & $0.7078_{\pm 0.0028}$ & $\underline{0.5116_{\pm 0.0040}}$ & $\underline{0.3628_{\pm 0.0019}}$ & 
            $0.7006_{\pm 0.0032}$ & $\underline{0.4204_{\pm 0.0012}}$ & $\underline{0.7232_{\pm 0.0014}}$ &  $0.6635_{\pm 0.0042}$ & $0.4876_{\pm 0.0066}$ & $\underline{0.8577_{\pm 0.0029}}$ & $0.5452_{\pm 0.0061}$ \\
            \cmidrule{2-12}
            & \multicolumn{1}{|c||}{\cellcolor{my_gray}DISCO} & \cellcolor{my_gray}$ \bm{0.7259^*_{\pm 0.0016}}$ & \cellcolor{my_gray}$\bm{0.5159^*_{\pm 0.0033}}$ & \cellcolor{my_gray}$\bm{0.3681^*_{\pm 0.0015}}$ & \cellcolor{my_gray}$\bm{0.7029^*_{\pm 0.0084}}$ &  \cellcolor{my_gray}$\bm{0.4266^*_{\pm 0.0029}}$ &
            \cellcolor{my_gray}$\bm{0.7274^*_{\pm 0.0056}}$ &  \cellcolor{my_gray}$\bm{0.6681^*_{\pm 0.0148}}$ & \cellcolor{my_gray}$\bm{0.4916^*_{\pm 0.0192}}$ & \cellcolor{my_gray}$\bm{0.8608^*_{\pm 0.0065}}$ & \cellcolor{my_gray}$\bm{0.5545^*_{\pm 0.0212}}$ \\
            \bottomrule
        \end{tabular}
    }
    \vspace{-4mm}
    
\end{table*}

\subsection{Experiment Setups}

\subsubsection{\textbf{Dataset Description and Preparation.}} 
In this paper, the real-world job recommendation datasets utilized for experiments were provided by an online recruitment platform. Specifically, it contains four kinds of behaviors: \textit{Browse}, \textit{Click}, \textit{Chat} and \textit{Match}. The behavior \textit{Match} is considered as the positive sample otherwise the negative sample (as mentioned in Section 3). To ensure reasonableness, we filtered out the job seekers with fewer than ten \textit{Match} interaction logs and jobs with fewer than five records. In particular, the dataset does not contain any sensitive information, and all IDs have been remapped by the provider to ensure they do not correspond to the original identifiers from the platform. Since the size of the entire interaction data is extremely tremendous, we selected two subsets based on the taxonomy of career clusters for our experiments: technology and service. In addition, to verify the availability and generalization of our model for other recommendation tasks, we also specifically selected a public educational dataset Edu-Rec~\cite{NIPS-Edu} for our experiments. For each dataset, we randomly split the job seeker-job position interaction data into three parts in the ratio of 7:1:2, serving as the training set, validation set, and testing set, respectively. The detailed statistical information is shown in~Table~I.





\begin{table}[!t]
    \setlength{\tabcolsep}{1.5pt}
    \renewcommand{\arraystretch}{1.18}
    \centering
    \label{table.Result_Table_2}
    \caption{Performance of DISCO and baselines on the Edu-Rec dataset. “$*$” denotes the statistically significant improvement where \textit{p}$<$0.05. $\bold{Bold}$: the best, $\underline{\rm Underline}$: the runner-up.}
    \vspace{-1mm}
    \resizebox{0.98\linewidth}{!}{ 
        \Huge
        \begin{tabular}{cc|c|c|c|c|c}
            \toprule
            \multicolumn{2}{c||}{\textbf{Datasets}}  & \multicolumn{5}{c}{\textbf{Edu-Rec}} \\
            \midrule
            \makecell[c]{\textbf{Base}\\ \textbf{Model}} & \multicolumn{1}{|c||}{\textbf{Method}}  & \textbf{AUC} & \textbf{HR@5} & \textbf{NDCG@5} & \textbf{HR@10} & \textbf{NDCG@10} \\
            \midrule
            \multirow{5}{*}{MF} & \multicolumn{1}{|c||}{Normal} &  $0.5805_{\pm 0.0014}$ & $0.1847_{\pm 0.0126}$ & $0.1210_{\pm 0.0162}$ & $0.3849_{\pm 0.0032}$ &  $0.1846_{\pm 0.0207}$ \\
            & \multicolumn{1}{|c||}{NCF} & $0.7085_{\pm 0.0004}$ &
            $0.2458_{\pm 0.0028}$ & $0.1680_{\pm 0.0025}$ & $0.4654_{\pm 0.0033}$ & $0.2381_{\pm 0.0023}$ \\
            & \multicolumn{1}{|c||}{AutoInt} & $0.7385_{\pm 0.0060}$ & $\underline{0.2644_{\pm 0.0017}}$ & $0.1709_{\pm 0.0045}$ & $0.4809_{\pm 0.0032}$ &  $0.2401_{\pm 0.0057}$ \\
            & \multicolumn{1}{|c||}{FINAL} & $\underline{0.7421_{\pm 0.0010}}$ & $0.2616_{\pm 0.0038}$ & $\underline{0.1799_{\pm 0.0010}}$ & $\underline{0.4844_{\pm 0.0027}}$ &
            $\underline{0.2509_{\pm 0.0013}}$ \\
            \cmidrule{2-7}
             & \multicolumn{1}{|c||}{\cellcolor{my_gray}DISCO}  & \cellcolor{my_gray}$ \bm{0.7519^*_{\pm 0.0008}}$ & \cellcolor{my_gray}$\bm{0.2704^*_{\pm 0.0015}}$ & \cellcolor{my_gray}$\bm{0.1871^*_{\pm 0.0010}}$ & \cellcolor{my_gray}$\bm{0.4988^*_{\pm 0.0006}}$ &
             \cellcolor{my_gray}$\bm{0.2582^*_{\pm 0.0012}}$ \\
             \midrule
             \multirow{5}{*}{NGCF} & \multicolumn{1}{|c||}{Normal} &  $\underline{0.7502_{\pm 0.0149}}$ & $\underline{0.2680_{\pm 0.0085}}$ & $0.1662_{\pm 0.0062}$ & $0.4959_{\pm 0.0055}$ &  $0.2389_{\pm 0.0053}$ \\
            & \multicolumn{1}{|c||}{NCF} & $0.7474_{\pm 0.0009}$ & $0.2565_{\pm 0.0033}$ & $0.1793_{\pm 0.0007}$ & $0.4872_{\pm 0.0033}$ &  $0.2527_{\pm 0.0006}$ \\
            & \multicolumn{1}{|c||}{AutoInt} & $0.7501_{\pm 0.0005}$ & $0.2655_{\pm 0.0029}$ & $0.1800_{\pm 0.0006}$ & $\underline{0.5007_{\pm 0.0026}}$ &
            $\underline{0.2549_{\pm 0.0006}}$ \\
            & \multicolumn{1}{|c||}{FINAL} &
            $0.7465_{\pm 0.0011}$ & $0.2528_{\pm 0.0025}$ & $\underline{0.1806_{\pm 0.0007}}$ & $0.4842_{\pm 0.0028}$ &
             $0.2540_{\pm 0.0010}$ \\
            \cmidrule{2-7}
             & \multicolumn{1}{|c||}{\cellcolor{my_gray}DISCO} & \cellcolor{my_gray}$ \bm{0.7641^*_{\pm 0.0001}}$ & \cellcolor{my_gray}$\bm{0.2778^*_{\pm 0.0011}}$ & \cellcolor{my_gray}$\bm{0.1887^*_{\pm 0.0008}}$ & \cellcolor{my_gray}$\bm{0.5033^*_{\pm 0.0014}}$ &
             \cellcolor{my_gray}$\bm{0.2615^*_{\pm 0.0005}}$ \\
             \midrule
             \multirow{5}{*}{LightGCN} & \multicolumn{1}{|c||}{Normal} &  $0.7422_{\pm 0.0013}$ & $0.2393_{\pm 0.0038}$ & $0.1796_{\pm 0.0020}$ & $0.4644_{\pm 0.0043}$ &  $0.2509_{\pm 0.0018}$ \\
            & \multicolumn{1}{|c||}{NCF} & $0.7456_{\pm 0.0005}$ & $0.2467_{\pm 0.0015}$ & $0.1799_{\pm 0.0008}$ & $0.4755_{\pm 0.0023}$ &  $0.2522_{\pm 0.0010}$ \\
            & \multicolumn{1}{|c||}{AutoInt} & $0.7544_{\pm 0.0006}$ & $0.2642_{\pm 0.0014}$ & $\underline{0.2026_{\pm 0.0039}}$ & $\underline{0.4940_{\pm 0.0012}}$ &
            $\underline{0.2761_{\pm 0.0045}}$ \\
            & \multicolumn{1}{|c||}{FINAL} & $\underline{0.7573_{\pm 0.0006}}$ &
            $\underline{0.2672_{\pm 0.0013}}$ & $0.1786_{\pm 0.0031}$ & $0.4939_{\pm 0.0011}$ & $0.2511_{\pm 0.0037}$ \\
            \cmidrule{2-7}
             & \multicolumn{1}{|c||}{\cellcolor{my_gray}DISCO} & \cellcolor{my_gray}$ \bm{0.7657^*_{\pm 0.0008}}$ & \cellcolor{my_gray}$\bm{0.2764^*_{\pm 0.0011}}$ & \cellcolor{my_gray}$\bm{0.2085^*_{\pm 0.0012}}$ & \cellcolor{my_gray}$\bm{0.5052^*_{\pm 0.0009}}$ &
             \cellcolor{my_gray}$\bm{0.2829^*_{\pm 0.0004}}$ \\
             \midrule
             \multirow{5}{*}{DPGNN} & \multicolumn{1}{|c||}{Normal} &  $0.7596_{\pm 0.0001}$ & $\underline{0.2883_{\pm 0.0015}}$ & $0.1820_{\pm 0.0009}$ & $0.5026_{\pm 0.0011}$ &  $0.2506_{\pm 0.0010}$ \\
            & \multicolumn{1}{|c||}{NCF} & $0.7600_{\pm 0.0003}$ & $0.2772_{\pm 0.0023}$ & $0.1807_{\pm 0.0012}$ & $0.5019_{\pm 0.0028}$ &  $0.2525_{\pm 0.0009}$ \\
            & \multicolumn{1}{|c||}{AutoInt} & $\underline{0.7602_{\pm 0.0001}}$ & $0.2803_{\pm 0.0015}$ & $0.1761_{\pm 0.0011}$ & $\underline{0.5029_{\pm 0.0018}}$ &
            $0.2472_{\pm 0.0016}$ \\
            & \multicolumn{1}{|c||}{FINAL} & $0.7572_{\pm 0.0003}$ &
            $0.2875_{\pm 0.0015}$ & $\underline{0.1872_{\pm 0.0014}}$ & $0.5016_{\pm 0.0012}$ & $\underline{0.2557_{\pm 0.0019}}$ \\
            \cmidrule{2-7}
             & \multicolumn{1}{|c||}{\cellcolor{my_gray}DISCO} & \cellcolor{my_gray}$ \bm{0.7676^*_{\pm 0.0008}}$ & \cellcolor{my_gray}$\bm{0.2945^*_{\pm 0.0010}}$ & \cellcolor{my_gray}$\bm{0.1949^*_{\pm 0.0011}}$ & \cellcolor{my_gray}$\bm{0.5063^*_{\pm 0.0006}}$ &
             \cellcolor{my_gray}$\bm{0.2618^*_{\pm 0.0010}}$ \\
            \bottomrule
        \end{tabular}
    }
    \vspace{-3mm}
    
\end{table}


\subsubsection{\textbf{Baseline Approaches.}}
Four widely used recommendation methods for basic representation learning are selected as the \textbf{base models} to obtain the representations of job seekers and job positions: \textbf{MF}~\cite{MF}, \textbf{NGCF}~\cite{NGCF}, \textbf{LightGCN}~\cite{LightGCN}, and \textbf{DPGNN}~\cite{DPGNN}. To evaluate the effectiveness of interaction modeling between job seekers and job positions, we incorporate three different \textbf{interaction modeling methods} into these base models, thereby constructing the complete baselines: \textbf{NCF}~\cite{NCF}, \textbf{AutoInt}~\cite{AutoInt} and \textbf{FINAL}~\cite{FINAL}. In addition, we also select two state-of-the-art methods \textbf{SHPJF}~\cite{SHPJF}, and \textbf{ECF}~\cite{ECF} for job recommendation and interpretable recommendation, respectively, to further validate the superiority of our DISCO.

\subsubsection{\textbf{Evaluation Protocols and Implementation Details.}}
To evaluate the performance of DISCO and all baseline methods, we employ three widely used metrics including the Area Under the ROC Curve (AUC), Hit Ratio (HR@$k$) and Normalized Discounted Cumulative Gain (NDCG@$k$). We empirically set $k$ to 5 and 10, and utilize these metrics to do the evaluation for the job recommendation task, i.e., predicting the matching probability and ranking jobs for candidates, which is more in line with the real online recruitment scenarios. Specifically, for each positive instance, we randomly sample 25 jobs for candidates as negative instances.

We implemented all models with Pytorch by Python and conducted our experiments on a Linux server with eight Nvidia A800 GPUs. We conducted each of the experiments 5 times and used the average value as the final result. The $t$-test was used to identify the significant differences between the performances of DISCO and the baselines. To perform the training process, we initialized all network parameters with Xavier initialization. Each parameter is sampled from $U\big(-\sqrt{2/(n_{in}+n_{out})}, \sqrt{2/(n_{in}+n_{out})}\big)$, where $n_{in}$ and $n_{out}$ denote the numbers of neurons feeding in and feeding out, respectively. The dimension size of hidden representations (i.e., $d$ and $d_h$) was set as 256. We use the Adam optimizer, where the learning rate was searched in \{5e-5, 8e-5, 1e-4, 2e-4, 5e-4\}. The coefficient $\lambda$ of contrastive loss was set to~1e-3.  

\subsection{Performance Comparison}
Table~II shows the experimental results of the proposed framework's job recommendation performance compared with the baselines on the two recruitment datasets. We highlighted the best results of all models in boldface and underlined the suboptimal results. According to the results, there are several observations: (1) Our DISCO framework embedded in the four underlying representational models presents significant advantages to all baselines. Specifically, our four models outperform the best baseline in terms of the AUC metric by an average of 0.65 and 0.64 on the two datasets, respectively. In particular, compared to the baseline based MF model, our model improves on average by 2.96 and 3.62 on both datasets with respect to the HR@5 and NDCG@5 metrics, respectively, which is considerably significant. (2) The models with the introduction of DISCO framework, show notably greater improvement in recommendation metrics than in classification metrics compared to all baselines, which further demonstrates our model's suitability for the job recommendation task. (3) In the baseline models, those based on the FINAL and AutoInt methods significantly outperform the others, proving that modeling high-order user-item interactions is effective in enhancing performance within the same representation learning mode. (4) Job recommendations using the two base models, NGCF and LightGCN, tend to be more effective than other model types. This may be attributed to the high-order connectivity of job seekers and jobs, which enhances the delivery of pertinent information and performance in real recruitment recommendation interaction data.

\begin{table}[!t]
    \huge
    \setlength{\tabcolsep}{1.5pt}
    \renewcommand{\arraystretch}{1.32}
    \centering
    \label{table.Result_Table_3}
    \caption{Performance of DISCO and recommendation methods on the Technology dataset. “$*$” denotes the statistically significant improvement where \textit{p}$<$0.05. $\bold{Bold}$: the best, $\underline{\rm Underline}$: the runner-up.}
    \vspace{-1mm}
    \resizebox{0.98\linewidth}{!}{ 
        \begin{tabular}{l|c|c|c|c|c}
            \toprule
            \textbf{Datasets}  & \multicolumn{5}{c}{\textbf{Technology}} \\
            \midrule
             \textbf{Rec Model}& \textbf{AUC} & \textbf{HR@5} & \textbf{NDCG@5} & \textbf{HR@10} & \textbf{NDCG@10} \\
            \midrule
            MF &  $0.6755_{\pm 0.0014}$ & $0.2518_{\pm 0.0135}$ & $0.2394_{\pm 0.0084}$ & $0.5835_{\pm 0.0121}$ &  $0.3428_{\pm 0.0088}$  \\
            NGCF & $0.7101_{\pm 0.0195}$ & $0.4705_{\pm 0.0268}$ & $0.3377_{\pm 0.0262}$ & $0.6144_{\pm 0.0472}$ & 
            $0.3840_{\pm 0.0205}$  \\
            LightGCN & $0.7085_{\pm 0.0204}$ & $0.4741_{\pm 0.0056}$ & $0.3112_{\pm 0.0322}$ & $0.6364_{\pm 0.0304}$ &  $0.3637_{\pm 0.0089}$ \\
            DPGNN & $0.7039_{\pm 0.0004}$ & $0.4986_{\pm 0.0025}$ & $0.3348_{\pm 0.0035}$ & $\underline{0.7018_{\pm 0.0019}}$ &  
            $0.4005_{\pm 0.0034}$ \\
            SHPJF & $0.7049_{\pm 0.0048}$ & $\underline{0.5194_{\pm 0.0140}}$ & $\underline{0.3623_{\pm 0.0110}}$ & $0.6965_{\pm 0.0133}$ &
            $\underline{0.4197_{\pm 0.0090}}$ \\

            ECF & $\underline{0.7120_{\pm 0.0013}}$ & $0.4029_{\pm 0.0043}$ & $0.2454_{\pm 0.0032}$ & $0.6488_{\pm 0.0052}$ &
            $0.3248_{\pm 0.0040}$ \\

            \midrule
            
            \rowcolor{my_gray} DISCO  & $ \bm{0.7408^*_{\pm 0.0013}}$ & $\bm{0.5311^*_{\pm 0.0016}}$ & $\bm{0.3739^*_{\pm 0.0031}}$ & $\bm{0.7145^*_{\pm 0.0042}}$ &
             $\bm{0.4338^*_{\pm 0.0041}}$ \\
            \bottomrule
        \end{tabular}
    }
    \vspace{-3mm}
    
\end{table}

\begin{figure}[!t] 
	\centering 
	\includegraphics[scale=0.49]{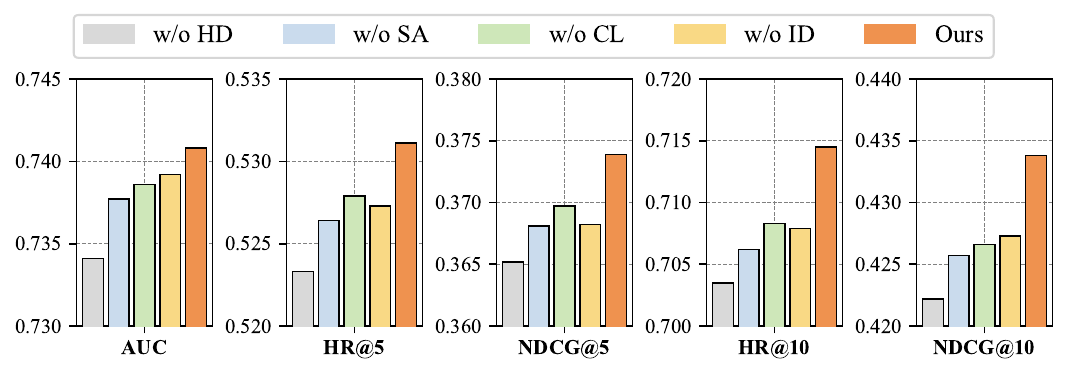} 
    \vspace{-3mm}
	\caption{Results of ablation study conducted on Technology dataset, where “w/o” means removing the target module.}
    \label{fig.case_study_1} 
    \vspace{-4mm}
\end{figure}

Meanwhile, Table III exhibits the experimental results on dataset Edu-Rec. It can be found that the proposed disco still holds a significant advantage over the compared interaction methods for the recommendation task on educational data, albeit with a drastic increase in the data size. This is further evidence that our approach is effective and generalizable for different recommendation tasks and not nearly limited to the domain of job recommendation. Furthermore, we also compared models specialized for job recommendation as well as interpretable recommendation models, respectively. As shown in Table IV, our DISCO has a relative improvement of 2.25\% and 3.20\% over the SHPJF model with respect to the recommended metrics HR@5 and NDCG@5, respectively. In particular, the relative improvement of our DISCO over the interpretable recommendation model ECF is 31.81\% and 52.36\% for these two metrics, respectively, which is relatively significant. These results strongly demonstrate the effectiveness and superiority of the proposed DISCO framework.

\vspace{-1mm}
\subsection{Ablation Study}
To answer RQ2, we conducted a comprehensive ablation experiment to investigate the effectiveness of each component in our DISCO framework. Specifically, we unfold the experiment with NGCF as the base model on the Technology dataset by defining the following variants: by defining the following variations: 1) \textbf{w/o HD}: removing the hierarchical skill-aware disentangling module and replacing it with a single granularity mapping; 2) \textbf{w/o SA}: removing the level-aware self-attention network; 3) \textbf{w/o CL}: removing the level-wise contrastive learning module; 4) \textbf{w/o ID}: removing the interaction diagnosis module and replace it with a normal prediction layer. As demonstrated in Figure~3, the results reveal insightful observations: (1) Compared to NGCF-DISCO, all variations experience a relative decline in performance in the Technology dataset in terms of various evaluation metrics, highlighting the significant impact of the designed submodules on our proposed framework. (2) The model performance shows the most considerable drop after eliminating the hierarchical skill-aware disentangling module, indicating that the idea of hierarchical disentangling works. This also validates the effectiveness of our designed model.

\begin{figure}[!t] 
	\centering 
	\includegraphics[scale=0.35]{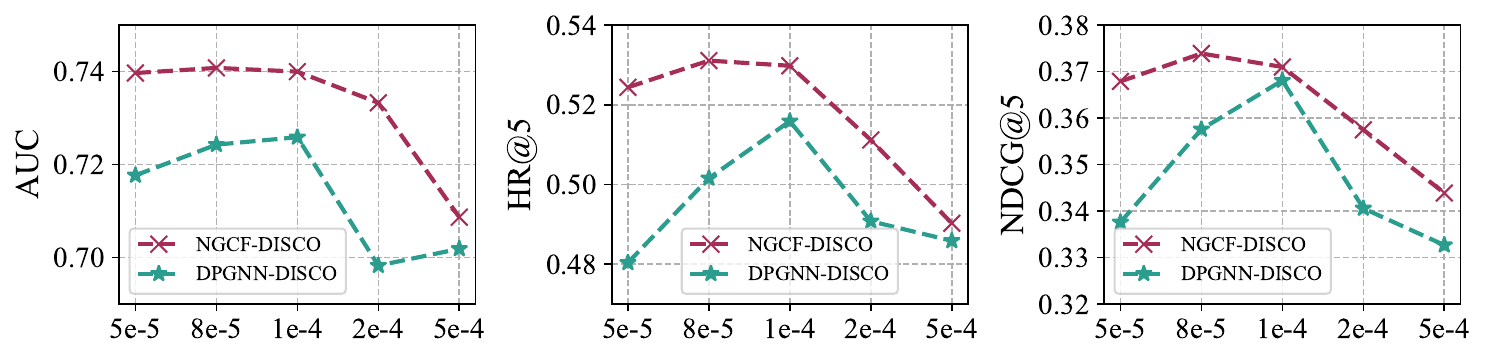} 
    \vspace{-3mm}
	\caption{Sensitivity analysis of learning rate for NGCF-DISCO and DPGNN-DISCO on Technology dataset.}
    \label{fig.para_study_1} 
    \vspace{-3mm}
\end{figure}

\begin{figure}[!t] 
    \vspace{-2mm}
	\centering 
	\includegraphics[scale=0.35]{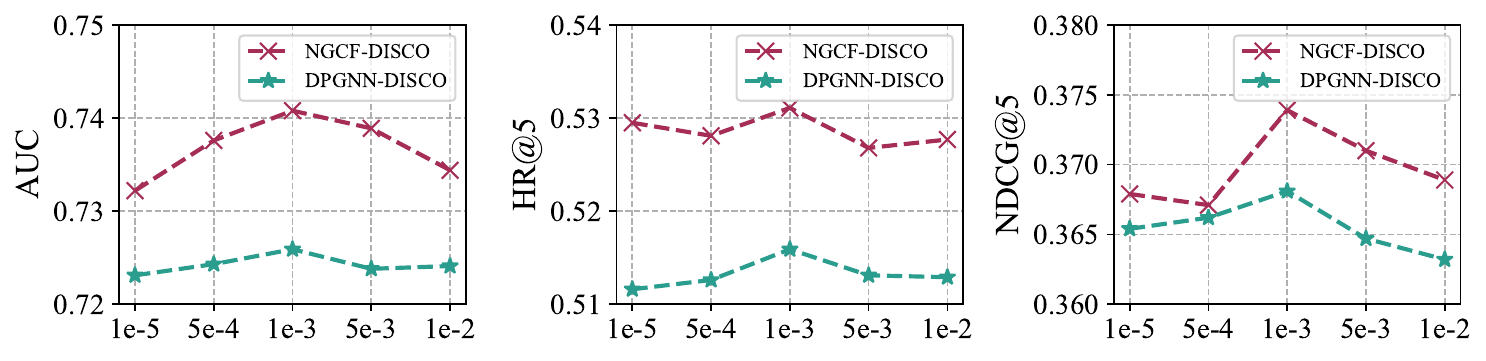} 
    \vspace{-3mm}
	\caption{Sensitivity analysis of coefficient $\lambda$ for NGCF-DISCO and DPGNN-DISCO on Technology dataset.}
    \label{fig.para_study_2} 
    \vspace{-3mm}
\end{figure}

\subsection{Parameter Sensitivity Analysis}
To answer RQ3, this section presents a parameter sensitivity analysis to explore the impacts of hyper-parameters, primarily focusing on the learning rate and the weight coefficient $\lambda$ of the contrastive loss. Specifically, we conduct experiments with NGCF and DPGNN as the base models on the Technology dataset, and set the list of learning rates to be \{5e-5, 8e-5, 1e-4, 2e-4, 5e-4\}, as well as the $\lambda$ values \{1e-5, 5e-4, 1e-3, 5e-3, 1e-2\}. As illustrated in Figure~4, We observe that different learning rates bring significantly different results. In particular, 8e-5 and 1e-4 are the optimal learning rates for NGCF-DISCO and DPGNN-DISCO, respectively, and both have a tendency to increase before decreasing. As shown in Figure~5, both models reach their best performance when the value of $\lambda$ is 1e-3. An interesting phenomenon here is that the trends of the coefficients affecting the performance are different with respect to the three metrics as the $\lambda$ value increments.

\subsection{Case Study}
To further explore the interpretability of our model, in particular the mining of job seekers' abilities and the difficulty of job skills, we conducted a case study. Specifically, we selected a pair of job seeker and position that achieved matching in the job search process and demonstrated interpretable content by outputting the hierarchical skill-associated representations from our model. As shown in Figure~6, it demonstrates candidate $c$'s mastery of each skill at the second and third levels (three level-2 skills and five level-3 skills are used here as examples, respectively), as well as the requirement values of the job for each skill. It can be observed that the candidate $c$'s skill proficiency at the fine-grained level influences the corresponding skill level at the coarse-grained level (e.g., his higher proficiency in $s_4$ and $s_5$ skews the level of $s_1$). Meanwhile, the candidate $c$'s proficiency level in each skill can be found to be generally compatible with the required level of the job $j$, which explains the pair's matching. The output from our model not only improves the interpretability of job recommendations, but also contributes to a deeper understanding of the job search process for both job seekers and recruiters.

\begin{figure}[!t] 
    \centering 
    \includegraphics[width=0.94\linewidth]{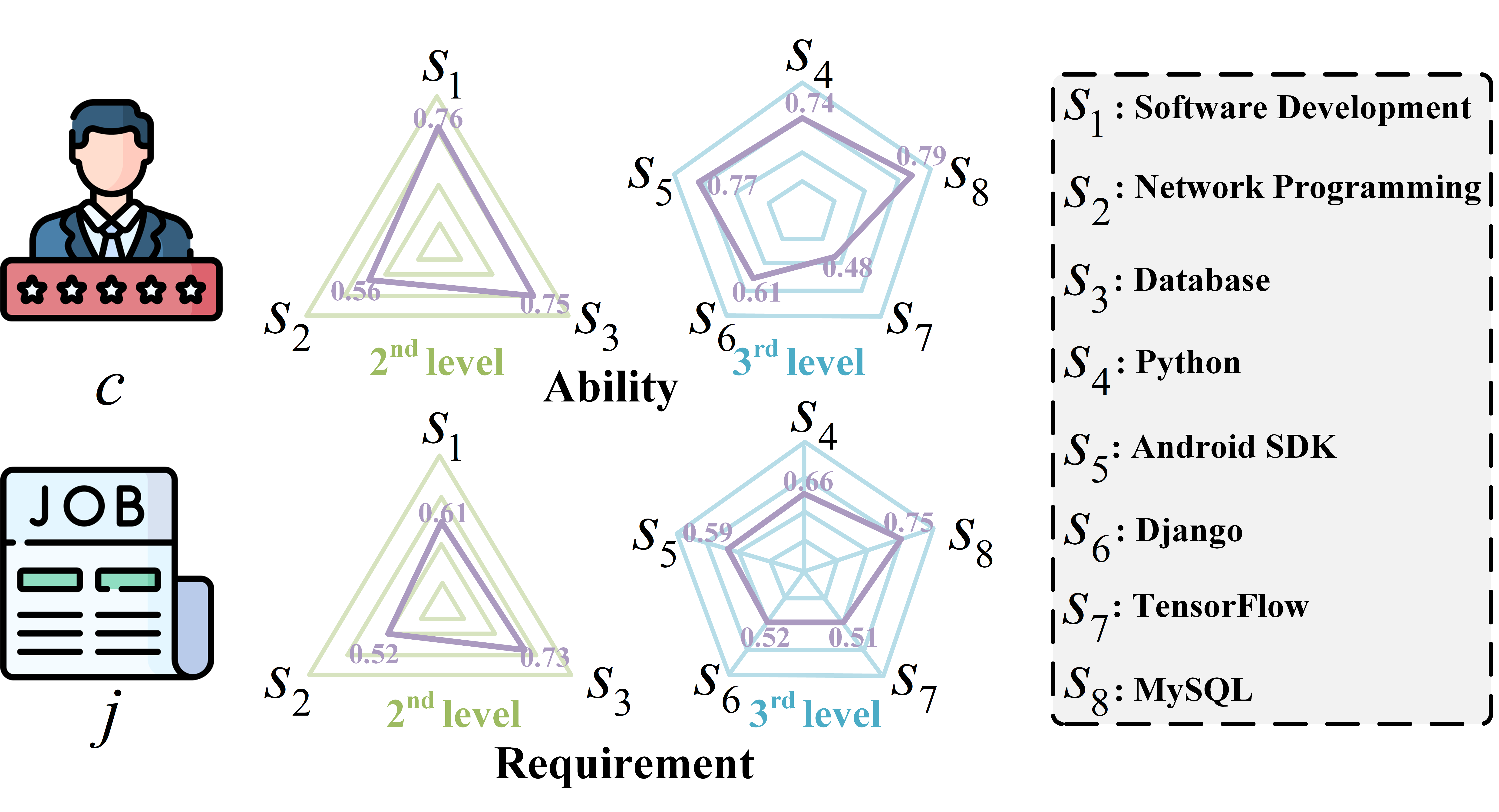} 
    \vspace{-2mm}
    \caption{Case study of the interpretability of our DISCO.}
    \label{fig.gcd} 
    \vspace{-6mm}
\end{figure}

\section{conclusion}

In this paper, we introduced a novel framework termed as \textbf{DISCO} (a hierarchical \underline{\textbf{Dis}}entangling based \underline{\textbf{Co}}gnitive diagnosis framework), which aims to flexibly accommodate the underlying representation learning model for job recommendations. Our approach comprises several key components. Initially, we designed a hierarchical representation disentangling module to mine the hierarchical skill-related factors embedded in the representations of job seekers and jobs. To further enhance information communication and robust representation learning, we proposed the level-aware association modeling, which consists of the inter-level knowledge influence module and level-wise contrastive learning. we devised an interaction diagnosis module is introduced that integrates a neural diagnosis function, aimed at effectively capturing the multi-level recruitment interaction process between job seekers and jobs. Finally, we developed an interaction diagnosis module incorporating a neural diagnosis function for effectively modeling the multi-level recruitment interaction process between job seekers and jobs, which introduces the cognitive measurement theory. Extensive experiments on two real-world recruitment recommendation datasets and an educational recommendation dataset clearly demonstrate the effectiveness and interpretability of our proposed DISCO framework. 

\section{Acknowledgments}
This work was supported in part by the National Natural Science Foundation of China (No.U21A20512, No.62302010, and No.62107001), in part by China Postdoctoral Science Foundation (No.2023M741849).

\bibliographystyle{IEEEtran}
\bibliography{references}

\end{document}